\documentstyle[prd,preprint,tighten,aps,eqsecnum,%
amssymb,newlfont,epsfig]{revtex}
\begin{document}
\draft
\preprint{
\begin{tabular}{r}
UWThPh-1999-41\\
DFTT 36/99\\
SISSA 89/99/EP\\
hep-ph/9907234\\
July 1999
\end{tabular}
}
\title{Constraints from Neutrino Oscillation Experiments 
on the Effective Majorana Mass in Neutrinoless Double $\beta$-Decay}
\author{S.M. Bilenky}
\address{Joint Institute for Nuclear Research, Dubna, Russia, and\\
Institute for Theoretical Physics, University of Vienna,\\
Boltzmanngasse 5, A--1090 Vienna, Austria}
\author{C. Giunti}
\address{INFN, Sezione di Torino, and Dipartimento di Fisica Teorica,
Universit\`a di Torino,\\
Via P. Giuria 1, I--10125 Torino, Italy}
\author{W. Grimus}
\address{Institute for Theoretical Physics, University of Vienna,\\
Boltzmanngasse 5, A--1090 Vienna, Austria}
\author{B. Kayser}
\address{National Science Foundation, Division of Physics,\\
Arlington, VA 22230, U.S.A.}
\author{S.T. Petcov\thanks{Also at: Institute of
Nuclear Research and Nuclear Energy, Bulgarian Academy of Sciences,
BG-1784, Sofia, Bulgaria}}
\address{Scuola Internazionale Superiore di Studi Avanzati,\\
and INFN, Sezione di Trieste, I-34013 Trieste, Italy}
\maketitle

\begin{abstract}
We determine the possible values of the effective Majorana neutrino mass 
$|\langle m \rangle |= |\sum_j U_{ej}^2 m_j|$ in the different
phenomenologically viable three and four-neutrino scenarios.
The quantities
$U_{\alpha j}$ ($\alpha = e,\mu,\tau,\ldots$) denote the elements of
the neutrino mixing matrix and the Majorana neutrino masses 
$m_j$ ($j=1,2,3,\ldots$) are ordered as 
$m_1 < m_2 < \ldots\,$ Assuming $m_1 \ll m_3$
in the three-neutrino case and $m_1 \ll m_4$ in the four-neutrino case,
we discuss, in particular, how constraints on $| \langle m \rangle |$ 
depend on the mixing angle
relevant in solar neutrino oscillations and on the three mass-squared
differences obtained from the analyses of
the solar, atmospheric and LSND data.
If neutrinoless double $\beta$-decay proceeds via the mechanism involving 
$|\langle m \rangle|$, conclusions about neutrinoless double $\beta$-decay can
be drawn. If one of the two viable four-neutrino schemes (Scheme A)
is realized in nature, $|\langle m \rangle|$ can be as large as 1 eV and
neutrinoless double 
$\beta$-decay could possibly be discovered in the near future. In this case
a Majorana CP phase of the mixing matrix $U$ could be determined.
In the other four-neutrino scheme (Scheme B) there is an upper bound 
on $|\langle m \rangle|$ of the order of $10^{-2}$ eV. 
In the case of three-neutrino mixing the same is true if the neutrino mass 
spectrum is hierarchical, however, 
if there exist two quasi-degenerate neutrinos and the first
neutrino has a much smaller mass, values of $|\langle m \rangle|$ 
as large as $\sim 0.1$ eV are possible.
\end{abstract}

\pacs{14.60.Pq, 14.60.St}     

\section{Introduction}

The observation of a significant up-down asymmetry of atmospheric multi-GeV
muon events by the Super-Kamiokande collaboration \cite{SK-atm} is considered
as a strong evidence in favour of neutrino masses and mixing.
The further investigation of neutrino properties and the understanding of the
origin of neutrino masses and mixing is a very important issue of present
day physics.
One of the most fundamental problems of the physics of neutrinos
is the question of
the nature of massive neutrinos: Are massive neutrinos
Dirac particles possessing some conserved lepton number or truly neutral
Majorana particles having all lepton numbers equal to zero?
Neutrino oscillation experiments cannot answer this question because the
additional phases of the neutrino mixing matrix in the Majorana case
do not enter into the transition probabilities
in vacuum
\cite{BHP80-Kobzarev80-Doi81}
as well as in matter
\cite{Langacker87}.

A direct way to reveal the nature of massive neutrinos is to
investigate processes in which the total lepton number is not conserved.
The most promising process of this type is neutrinoless double
$\beta$-decay of nuclei (for reviews see \cite{bb-review}),
\begin{equation}\label{bb}
(A,Z) \to (A,Z+2) + e^- + e^- \,.
\end{equation}

In the framework of the standard left-handed weak interactions
with the Hamiltonian for $\beta$-decay given by
\begin{equation}\label{H}
\mathcal{H}_\beta = \sqrt{2} G_F\, \bar e_L \gamma_\alpha \nu_{eL} \,
j^\alpha + \mbox{h.c.}\,,
\end{equation}
where $j^\alpha$ is the weak hadronic current,
and with the mixing of Majorana neutrinos
\begin{equation}\label{mixing}
\nu_{\alpha L} = \sum_j U_{\alpha j}\nu_{jL}
\quad \mbox{with} \quad \alpha = e, \mu, \tau, \ldots,
\end{equation}
where $\nu_j = \nu_j^c \equiv C \bar{\nu}_j^T$ ($j=1,2,3,\ldots$)
is the field of a Majorana neutrino with mass $m_j$,
the matrix element of the process (\ref{bb}) is proportional to the effective
Majorana neutrino mass
\begin{equation}\label{m}
|\langle m \rangle |= \left|\sum_j U^2_{ej} m_j \right| \,,
\end{equation}
which originates from the neutrino propagator
\begin{equation}\label{prop}
\langle 0 | T \left[ \nu_{eL}(x_1) \nu_{eL}^T(x_2) \right] | 0 \rangle \,.
\end{equation}

In this paper, assuming that massive neutrinos are Majorana particles, 
we will present constraints on the parameter
$|\langle m \rangle|$ 
that can be obtained from results of neutrino oscillation experiments
\cite{Petcov-Smirnov-94,BBGK,BGKP,BG-bb,Minakata-Yasuda-bb,%
Fukuyama-bb,Adhikari-Rajasekaran-98,%
BG-WIN99,barger99,Vissani-bb,carlo,petcov}
in
the framework of the two possible schemes of mixing of four massive
neutrinos that can accommodate the results of all existing
neutrino oscillation experiments (Schemes A and B \cite{BGG-AB})
and two schemes of mixing of three massive neutrinos
with $m_1 \ll m_3$ that can 
accommodate the results of all experiments except the 
results of the LSND experiment
(the scheme with a hierarchy of neutrino masses and the scheme with the 
reversed hierarchy of neutrino masses, i.e., with quasi-degenerate masses 
$m_2$, $m_3$)
It will be shown that in one of the four-neutrino mixing schemes
(Scheme A) the effective Majorana mass $|\langle m \rangle|$ 
can be as large as the ``LSND mass'' 
$\sqrt{\Delta m^2_\mathrm{LSND}}$ lying between 0.4 and 1.4 eV.
In the three-neutrino mixing scheme with reversed
mass hierarchy $|\langle m \rangle|$ can be equal to the
``atmospheric neutrino mass'' $\sqrt{\Delta m^2_\mathrm{atm}}$ ranging from
0.03 to 0.1 eV.
In two other schemes the effective Majorana mass is strongly suppressed
with respect to the masses of the heaviest neutrinos.
We will show that in the case of the three-neutrino mass hierarchy 
and in Scheme B of the mixing of four neutrinos
the upper bound on $|\langle m \rangle|$ is in the range of a few
$10^{-2}$ eV.

Many experiments on the search for $(\beta\beta)_{0\nu}$ decay of different
nuclei are going on at present (for a review see Ref. \cite{morales}). 
Up to now neutrinoless double $\beta$-decay has not been found, 
therefore, only 
lower bounds on the half-life of the $(\beta\beta)_{0\nu}$ decay modes
can be inferred from the experimental data. The most stringent limit 
was obtained in the $^{76}$Ge Heidelberg -- Moscow experiment \cite{baudis}. 
The latest result of this experiment is 
\begin{equation}
T_{1/2} > 5.7 \times 10^{25} \: \mathrm{y} \,.
\end{equation}
Concerning other nuclei, the best limit for $^{136}$Xe 
($T_{1/2} > 4.4 \times 10^{23} \: \mathrm{y}$) has been obtained by the 
Caltech -- PSI -- Neuch\^atel Coll.  \cite{caltech} and for
$^{130}$Te ($T_{1/2} > 7.7 \times 10^{22} \: \mathrm{y}$) by the
Milano Group \cite{milano}.
For a list of recent experimental results see Table 2 in
Ref. \cite{morales}. 

From the lower bounds on $T_{1/2}$ upper bounds on the effective Majorana
mass $| \langle m \rangle |$ can be inferred. Taking the result of the
$^{76}$Ge experiment, it was found (see references cited in \cite{baudis}) 
that
\begin{equation}
| \langle m \rangle | \lesssim 0.2 - 0.6 \: \mathrm{eV} \,.
\end{equation}
Such bounds are obtained by using the results of the
calculations of the nuclear matrix elements of the relevant isotopes,
performed by different groups in the framework of the Shell Model or 
the Quasiparticle Random Phase Approximation.
For a discussion of calculations of $(\beta\beta)_{0\nu}$ nuclear 
matrix elements see Ref. \cite{morales}.

Running experiments or experiments under preparation plan
to reach a sensitivity of 
$| \langle m \rangle | \sim 10^{-1}$ eV. The feasibility of a new generation
of $(\beta\beta)_{0\nu}$ decay experiments,
CUORE \cite{milano} and GENIUS \cite{GENIUS},
which could reach the region
$| \langle m \rangle | \sim 10^{-2}$ eV, is
under consideration at the moment.

The mechanism for 
$(\beta\beta)_{0\nu}$ decay based on the propagator (\ref{prop}) is not
unique. In addition to the Majorana neutrino exchange, 
there can be other possible mechanisms 
involving physics 
beyond the standard model: right-handed
currents, SUSY with violation of R-parity, mechanisms with scalars, 
etc. (for a review see Ref. \cite{Mohapatra-98}). 
It has been argued that, independently of the underlying 
mechanism, an observation of $(\beta\beta)_{0\nu}$ decay would be an
evidence for a non-zero \emph{Majorana} neutrino mass. In 
Ref. \cite{schechter} the argument is phrased in terms of Feynman graphs, 
whereas in Ref. \cite{takasugi} symmetry reasons are given.
However, these arguments do not allow to calculate
the Majorana neutrino mass. Actually, the contribution to the Majorana
neutrino mass suggested by these arguments is exceedingly small.
Thus, it is clear that any 
information on $|\langle m \rangle|$ that
can be obtained from neutrino oscillation data could help to reveal the
true mechanism of $(\beta\beta)_{0\nu}$ decay. 

\section{Neutrinoless double $\beta$-decay in four-neutrino schemes}

Results from many neutrino oscillation experiments are available at present.
Convincing evidence in favour of neutrino masses and
oscillations have been obtained in atmospheric
\cite{SK-atm,atm-exp} and solar neutrino
experiments \cite{sun-exp,SK-sun}. Observation of 
$
\stackrel{\scriptscriptstyle (-)}{\nu}_{\hskip-3pt \mu} 
\to
\stackrel{\scriptscriptstyle (-)}{\nu}_{\hskip-3pt e}
$
oscillations has been claimed by the LSND collaboration \cite{LSND}.  
On the other hand, in different reactor and accelerator short-baseline
(SBL) experiments (see \cite{SBL-exp}) and in the long-baseline (LBL) 
reactor experiment CHOOZ \cite{CHOOZ}
no indications in favour of neutrino oscillations were found.

From the analysis of all these data it follows that there are three
different scales of neutrino mass-squared differences $\Delta m^2$:
\begin{equation}\label{scales}
\Delta m^2_\mathrm{solar} \sim 10^{-5} \; \mathrm{eV}^2 \; \mathrm{or} \;
10^{-10} \; \mathrm{eV}^2, \quad \Delta m^2_\mathrm{atm} \sim 10^{-3} \;
\mathrm{eV}^2, \quad \Delta m^2_\mathrm{LSND} \sim 1 \; \mathrm{eV}^2 \,.
\end{equation}
In order to
accommodate all these data we need to assume that at least four massive
neutrinos exist in nature. This means that, in addition to the three flavour
neutrinos, sterile neutrinos must exist. 

In the framework of the minimal
scheme with four massive neutrinos
\cite{four-models,four-phenomenology,BBN,BGKP,BGG-AB},
it was shown in Ref. \cite{BGG-AB} 
that from the
six possible types of mass spectra with mass-squared differences
(\ref{scales}) only two are compatible with all data, including the latest
Super-Kamiokande data on the measurement of the up-down asymmetry of multi-GeV
atmospheric muon events:
\begin{equation} \label{AB}
\mbox{(A)}
\qquad
\underbrace{
\overbrace{m_1 < m_2}^{\mathrm{atm}}
\ll
\overbrace{m_3 < m_4}^{\mathrm{solar}}
}_{\mathrm{LSND}}
\qquad \mbox{and} \qquad
\mbox{(B)}
\qquad
\underbrace{
\overbrace{m_1 < m_2}^{\mathrm{solar}}
\ll
\overbrace{m_3 < m_4}^{\mathrm{atm}}
}_{\mathrm{LSND}}
\;.
\end{equation}

In order to determine the effective Majorana mass $| \langle m \rangle |$
(\ref{m}) we need to know the neutrino masses and
the elements $U_{ej}$ of the neutrino mixing matrix.
Information on these elements can be obtained from the results
of reactor and solar neutrino experiments. In the framework of Schemes A
and B the $\bar \nu_e$ survival probability in SBL experiments
is given by \cite{BGG-AB}
\begin{equation}
P^\mathrm{SBL}_{\bar\nu_e\to\bar\nu_e} = 1 - \frac{1}{2} B_{e;e}
\left( 1 - \cos \frac{\Delta m^2_{41}L}{2p} \right) \,,
\end{equation}
where $\Delta m^2_{41} = m_4^2-m_1^2$ is the largest neutrino mass-squared
difference, $L$ is the source -- detector distance, $p$ denotes the neutrino
momentum and 
\begin{equation}
B_{e;e} = 4 \sum_{j=3,4} |U_{ej}|^2 \left( 1-\sum_{j=3,4} |U_{ej}|^2 \right)
\end{equation}
represents the oscillation amplitude. Taking into account the results of the
solar neutrino experiments, the negative results of the reactor experiments
allow to deduce the bounds
\begin{equation}\label{A}
\sum_{j=3,4} |U_{ej}|^2 \geq 1-a^0_e \quad \mbox{in Scheme A} 
\end{equation}
and
\begin{equation}\label{B}
\sum_{j=3,4} |U_{ej}|^2 \leq a^0_e \quad \mbox{in Scheme B} \,.
\end{equation}
The quantity $a^0_e$ is defined by
\begin{equation}\label{a0}
a^0_e = \frac{1}{2} \left( 1 - \sqrt{1-B^0_{e;e}} \right) \,,
\end{equation}
where $B^0_{e;e}$ is the upper bound on the amplitude
$B_{e;e}$. For each value of the mass-squared difference $\Delta m^2_{41}$, the value of
$B^0_{e;e}$ is obtained from the exclusion curves in the plane of oscillation
parameters obtained in $\bar\nu_e$ disappearance experiments.
Consequently, $B^0_{e;e}$ and $a^0_e$ are  
functions of $\Delta m^2_{41}$. 
Using the result of the Bugey experiment \cite{Bugey} we have 
$a^0_e \lesssim 4 \times 10^{-2}$ for $\Delta m^2_{41} \gtrsim 0.1$ 
eV$^2$. Now we will consider the effective Majorana mass 
$|\langle m \rangle|$ in Schemes A and B.

\subsection{Scheme A}
\label{4nuA}

With the assumption $m_{1,2} \ll m_{3,4}$ and taking into account the heaviest
neutrino masses, we get
\begin{equation}
|\langle m \rangle | \simeq | \sum_{j=3,4} U^2_{ej} |\, m_4 \,,
\end{equation}
where 
$m_4 \simeq \sqrt{\Delta m^2_\mathrm{LSND}}$.  
From Eq.(\ref{A}) it follows that, up to corrections of order $10^{-2}$,
the relation
\begin{equation}
\sum_{j=3,4} |U_{ej}|^2 \simeq 1
\end{equation}
is valid. This allows the approximate parameterization
\begin{equation}\label{param}
U_{e3} \simeq \cos \theta\, e^{i\alpha_3}, \quad 
U_{e4} \simeq \sin \theta\, e^{i\alpha_4} 
\end{equation}
and, therefore, \cite{BGKP}
\begin{equation}\label{Aeq}
|\langle m \rangle | \simeq 
\sqrt{1 - \sin^2 2\theta \sin^2 \alpha}\: m_4 \,,
\end{equation}
where $\alpha = \alpha_4 - \alpha_3$. Note that $\alpha$ does not
have any effect in neutrinos oscillations, i.e., it is one of the additional
phases in the Majorana case.

If CP is conserved in the lepton sector and if we assume a trivial CP phase
for the transformation of the electron under CP,\footnote{A CP phase
of the electron has no impact on our discussion.} we have the CP 
transformation
\begin{equation}\label{CPt}
e (x) \to -C e^* (x^0, -\vec x)\,, \quad
\nu_j (x) \to \eta_j C \nu_j^* (x^0, -\vec x)\,,
\end{equation}
where $C$ is the charge conjugation matrix. 
Invariance of the Majorana mass term 
$\frac{1}{2} m_j \nu_j^T C^{-1} \nu_j$ under this CP transformation 
leads to the condition
\begin{equation}\label{eta}
\eta_j = i \rho_j = \pm i
\end{equation}
for the phases $\eta_j$, which are called
CP parities of the Majorana fields $\nu_j$.
Thus, with Eqs.(\ref{CPt}) and (\ref{eta}) we obtain for Majorana 
neutrinos 
\begin{equation}\label{CP}
U_{ej} = U_{ej}^* \eta_j \,.
\end{equation}
From Eq.(\ref{CP}) it follows that
\begin{equation}
\alpha_j = \frac{\pi}{4} \rho_j + n_j \pi \,,
\end{equation}
where the $n_j$ are integers. Therefore, in the case of CP conservation, 
$\sin^2 \alpha = 0$ for equal CP parities and $\sin^2 \alpha = 1$ for opposite
CP parities.

From Eq.(\ref{Aeq}) it follows that
\begin{equation}\label{Aineq}
\sqrt{1 - \sin^2 2\theta} \sqrt{\Delta m^2_\mathrm{LSND}} 
\lesssim | \langle m \rangle | \lesssim \sqrt{\Delta m^2_\mathrm{LSND}} \,,
\end{equation}
where the boundary values correspond to CP conservation. The angle $\theta$
can be determined from the analysis of
the results of solar neutrino experiments.
In fact, the survival probability of solar $\nu_e$'s 
in Scheme A is given by \cite{BGKP}
\begin{equation}
P^\mathrm{solar}_{\nu_e\to\nu_e} = 
\left( 1 - \sum_{j=1,2} |U_{ej}|^2 \right)^2 
P_{\nu_e\to\nu_e} ( \sin^2 2\theta_\mathrm{solar}, \Delta m^2_\mathrm{solar} )
+ \sum_{j=1,2} |U_{ej}|^4 \,.
\end{equation}
Here $P_{\nu_e\to\nu_e} ( \sin^2 2\theta_\mathrm{solar}, 
\Delta m^2_\mathrm{solar} )$ 
is the standard two-neutrino $\nu_e$ survival probability with
\begin{eqnarray}
&& \cos^2 \theta_\mathrm{solar} = \frac{|U_{e3}|^2}{1-\sum_{j=1,2} |U_{ej}|^2}
   \simeq |U_{e3}|^2 \,, \nonumber \\
&& \sin^2 \theta_\mathrm{solar} = \frac{|U_{e4}|^2}{1-\sum_{j=1,2} |U_{ej}|^2}
   \simeq |U_{e4}|^2 \,. \label{theta}
\end{eqnarray}
Comparing Eqs.(\ref{param}) and (\ref{theta}) we conclude that
\begin{equation}
\sin^2 2\theta = \sin^2 2\theta_\mathrm{solar} \,.
\end{equation}

It is well known that from the analysis of solar neutrino data
two matter MSW solutions and one vacuum oscillation (VO)
solution of the solar neutrino problem have been found
(see the recent analyses in \cite{BKS98,sun-analysis,GG99}
and references therein).
To get 
an idea of the values of $\sin^2 2\theta_\mathrm{solar}$,
we quote the best-fit values of the combined analysis of 
Ref. \cite{GG99} for the MSW solutions and for the VO solution the 
best fit-value of Ref. \cite{BKS98}, which takes 
into account the event rates measured in the solar neutrino experiments:
\begin{enumerate}
\item the small mixing angle MSW solution (SMA) with 
$\sin^2 2\theta_\mathrm{solar} = 4.5 \times 10^{-3}$ and
$\Delta m^2_\mathrm{solar} = 6.3 \times 10^{-6}$ eV$^2$ for
transitions of solar $\nu_e$'s into active neutrinos and
$\sin^2 2\theta_\mathrm{solar} = 3.2 \times 10^{-3}$ and
$\Delta m^2_\mathrm{solar} = 5.0 \times 10^{-6}$ eV$^2$ for
transitions into sterile neutrinos,
\item the large mixing angle MSW solution (LMA) with
$\sin^2 2\theta_\mathrm{solar} = 0.80$ and
$\Delta m^2_\mathrm{solar} = 3.6 \times 10^{-5}$ eV$^2$,
\item the vacuum oscillation solution (VO) with
$\sin^2 2\theta_\mathrm{solar} = 0.75$ and
$\Delta m^2_\mathrm{solar} =  8.0 \times 10^{-11}$ eV$^2$.
\end{enumerate}
Note that for transitions into sterile neutrinos the data allow only
the SMA MSW solution. From Eq.(\ref{Aineq}) 
for the SMA MSW solution we get the relation
\begin{equation}
| \langle m \rangle | \simeq \sqrt{\Delta m^2_\mathrm{LSND}} \,,
\end{equation}
which is independent of CP violation. 
If this possibility is realized in nature,
$(\beta\beta)_{0\nu}$ experiments are expected to see an effect in the near
future if $\sqrt{\Delta m^2_\mathrm{LSND}} \gtrsim 0.5$ eV. 
If, however, it will be found that $| \langle m \rangle | \ll 0.5$ eV,
within Scheme A with Majorana neutrinos
the SMA MSW solution of the solar neutrino problem 
will be incompatible with the data. Thus, an
experimental confirmation of the SMA MSW solution 
of the solar neutrino problem and the indicated stringent
upper limit on $| \langle m \rangle |$ would allow to rule out Scheme
A with massive Majorana neutrinos.

It was shown in Ref. \cite{BBN} that,
if the effective number $N_\nu$ of neutrinos 
relevant in big bang nucleosynthesis 
is smaller than 4, then in both Schemes A and B the solution to 
the solar neutrino puzzle in the framework of neutrino oscillations 
is given by transitions into sterile neutrinos and consequently by
the SMA MSW solution. If $N_\nu < 4$ is not correct, $\nu_e$ could
in principle make transitions into an arbitrary admixture of 
$\nu_\tau$ and $\nu_s$ in Schemes A and B.
However, for the LMA MSW and VO solutions
an analysis of the solar neutrino data should certainly put a constraint
on the admixture of sterile neutrinos.

With increasing accuracy of 
the Super-Kamiokande measurements of the day-night asymmetry
and the electron recoil energy spectrum and with future results
of the SNO experiment \cite{SNO} the
preferred solution of the solar neutrino problem can possibly be found 
\cite{SK-sun}. 
In the case of the LMA MSW or the VO solution,
assuming $| \langle m \rangle |$ and $\sin^2
2\theta_\mathrm{solar}$ to be measured, we arrive at
\begin{equation}
\sin^2 \alpha \simeq \frac{1}{\sin^2 2\theta_\mathrm{solar}}
\left( 1 - \frac{| \langle m \rangle |^2}{\Delta m^2_\mathrm{LSND}} \right)
\end{equation}
from Eq.(\ref{Aeq}). Thus, if future measurements 
show the correctness of a large mixing angle solution of the solar 
neutrino problem (VO or MSW), then from measurements of 
$| \langle m \rangle |$ information on the
Majorana phase $\alpha$ of the mixing matrix $U$ can be obtained.

Finally, in Scheme A the neutrino mass 
which is probed in $^3$H $\beta$-decay spectrum is given by
\begin{equation}\label{H3}
m(^3\mbox{H}) \simeq m_4 \simeq \sqrt{\Delta m^2_\mathrm{LSND}}\,.
\end{equation}
For $\sqrt{\Delta m^2_\mathrm{LSND}} \sim 1$ eV,
$m(^3\mbox{H})$ lies in the 
sensitivity region of future $^3$H experiments \cite{tritium-future}. 
The check of the relation (\ref{H3}) is an additional test of Scheme A.

\subsection{Scheme B}

In Scheme B we have 
\begin{equation}
\sum_{j=3,4} |U_{ej}|^2 \leq a^0_e \,.
\end{equation}
Thus the contribution of the heavy masses 
$m_3 \simeq m_4 \simeq \sqrt{\Delta m^2_\mathrm{LSND}}$ to the effective
Majorana mass $| \langle m \rangle |$ is suppressed in this scheme.
Taking into account also the light masses $m_{1,2}$, we obtain the bound
\begin{equation}
| \langle m \rangle | \lesssim 
| \langle m \rangle |_2 + | \langle m \rangle |_{34} \,,
\end{equation}
where
\begin{equation}\label{b34}
| \langle m \rangle |_{34} = 
a^0_e \sqrt{\Delta m^2_\mathrm{LSND}}
\end{equation}
is the upper bound of the contribution of $\nu_{3,4}$ to 
$| \langle m \rangle |$ and, with the assumption
$m_1 \ll \sqrt{\Delta m^2_\mathrm{solar}}$,
\begin{equation}\label{b2}
| \langle m \rangle |_2 = 
\sin^2 \theta_\mathrm{solar} \sqrt{\Delta m^2_\mathrm{solar}}
\end{equation}
is the contribution of $\nu_2$ with
\begin{equation}\label{sin2}
\sin^2 \theta_\mathrm{solar} =
\frac{1}{2} \left( 1 - \sqrt{1 - \sin^2 2\theta_\mathrm{solar}} \right)\,.
\end{equation}

Note that in the above consideration we have neglected $m_1$. 
If $m_1$ and $m_2$ are of 
the same order of magnitude, we have
\begin{equation}\label{b2m}
| \langle m \rangle |_2 = 
m_1 +
\sin^2 \theta_\mathrm{solar} 
\frac{\Delta m^2_\mathrm{solar}}{m_1 + \sqrt{m^2_1 +
\Delta m^2_\mathrm{solar}}} \,,
\end{equation}
which reduces to Eq.(\ref{b2}) for 
$m_1 \ll \sin^2 \theta_\mathrm{solar} 
\sqrt{\Delta m^2_\mathrm{solar}}$. As long as
$m_1 \lesssim \sqrt{\Delta m^2_\mathrm{solar}}$ holds, the numerical value of 
the second term of the bound $| \langle m \rangle |_2$ 
does not change very much as a function of $m_1$.

\section{Neutrinoless double $\beta$-decay in three-neutrino schemes}

If the LSND indications in favour of 
$
\stackrel{\scriptscriptstyle (-)}{\nu}_{\hskip-3pt \mu} 
\to
\stackrel{\scriptscriptstyle (-)}{\nu}_{\hskip-3pt e}
$
oscillations are not confirmed by future experiments it is sufficient 
to assume the
existence of only three light massive neutrinos. With
$m_1 \ll m_3$ there are two possible
mass spectra in this case: 
\begin{enumerate}
\renewcommand{\labelenumi}{\Roman{enumi}.}
\item
\emph{The hierarchical spectrum} defined by\footnote{For the calculation
of $| \langle m \rangle |$ we will also allow for
$m_1 \sim m_2$. \label{I}}
$m_1 \ll m_2 \ll m_3$ with
$\Delta m^2_{21} = \Delta m^2_\mathrm{solar}$ and
$\Delta m^2_{31} = \Delta m^2_\mathrm{atm}$.
\item
\emph{The spectrum with reversed hierarchy} defined by 
$m_3 > m_2 \gg m_1$ with 
$\Delta m^2_{32} = \Delta m^2_\mathrm{solar}$ and 
$\Delta m^2_{31} = \Delta m^2_\mathrm{atm}$.
In this case $m_2$ and $m_3$ are quasi-degenerate.
\end{enumerate}
For a discussion of $| \langle m \rangle |$ with large $m_1$, 
allowing thus for degeneracy of the three neutrino masses,
see Refs. \cite{Petcov-Smirnov-94,barger99,Vissani-bb}.

\subsection{Neutrino mass hierarchy}
\label{3I}

In the case of the hierarchical neutrino mass spectrum 
(see also footnote \ref{I}) we have the
upper bound
\begin{equation}
| \langle m \rangle | \lesssim 
| \langle m \rangle |_2 + | \langle m \rangle |_3 \,,
\end{equation}
where
\begin{equation}\label{b3}
| \langle m \rangle |_3 = 
|U_{e3}|^2 \sqrt{\Delta m^2_\mathrm{atm}}
\end{equation}
and $| \langle m \rangle |_2$ is given by Eq.(\ref{b2m}).\footnote{If
we include the effect of non-zero $U_{e3}$, then we have
$$
| \langle m \rangle |_2 = (1 - |U_{e3}|^2) \left(
m_1 +
\sin^2 \theta_\mathrm{solar} 
\frac{\Delta m^2_\mathrm{solar}}{m_1 + \sqrt{m^2_1 +
\Delta m^2_\mathrm{solar}}} \right).
$$}

From the result of the LBL reactor experiment CHOOZ \cite{CHOOZ}, 
if one takes into account that $|U_{e3}|^2$ cannot be
large because of the results of the solar neutrino experiments,
one can find an upper bound on $|U_{e3}|^2$. With
the upper bound $B^\mathrm{CHOOZ}_{e;e}$ on the amplitude of
$\bar\nu_e\to\bar\nu_e$ oscillations we have 
\begin{equation}
|U_{e3}|^2 \leq \frac{1}{2} \left(
1-\sqrt{1-B^\mathrm{CHOOZ}_{e;e}} \right) \,,
\end{equation}
and from the 90\% CL CHOOZ exclusion plot one obtains
$|U_{e3}|^2 \lesssim 5 \times 10^{-2}$ for 
$\Delta m^2_{31} = \Delta m^2_\mathrm{atm} \gtrsim 2 \times 10^{-3}$
eV$^2$. From a 3-neutrino analysis of the Super-Kamiokande + CHOOZ data, 
in Ref. \cite{FLMS-98} it was found that 
$|U_{e3}|^2 \lesssim 0.15$ at 90\% CL, valid in the
whole range of $\Delta m^2_\mathrm{atm}$. A similar result
has been derived in Ref. \cite{barger98} 
($|U_{e3}|^2 \lesssim 0.1$ at 95\% CL). These bounds on $|U_{e3}|^2$
allow us to evaluate $| \langle m \rangle |_3$ (\ref{b3})
(see summary).

\subsection{The reversed hierarchy}

If this neutrino mass spectrum is realized in nature then the
bounds on $|U_{e3}|^2$ mentioned before are now valid for
$|U_{e1}|^2$. Thus the contribution of $m_1$ to 
$|\langle m \rangle|$ (\ref{m}) can
be neglected and we obtain
\begin{equation}
| \langle m \rangle | \simeq 
\left| \sum_{j=2,3} U^2_{ej} \right| \sqrt{\Delta m^2_{31}} 
\end{equation}
with $\Delta m^2_{31} = \Delta m^2_\mathrm{atm}$. Taking into account
that $\sum_{j=2,3} |U_{ej}|^2$ is close to 1 and that 
$\Delta m^2_{32} = \Delta m^2_\mathrm{solar}$, then with the methods of
Subsection \ref{4nuA} we arrive at the expression \cite{BGKP} 
\begin{equation}\label{3eqq}
|\langle m \rangle | \simeq 
\sqrt{1 - \sin^2 2\theta_\mathrm{solar} 
\sin^2 \alpha}\: \sqrt{\Delta m^2_\mathrm{atm}} \,,
\end{equation}
where now $\alpha = \alpha_3 - \alpha_2$ and $\alpha_{2,3}$ are the
phases of $U_{e2}$ and $U_{e3}$, respectively.
The discussion in Subsection \ref{4nuA} on the possibility of the 
determination of the CP phase $\alpha$ is also applicable here
(for further details see \cite{BGKP}).
From Eq.(\ref{3eqq}) we get the bounds
\begin{equation}\label{3ineq}
\sqrt{1 - \sin^2 2\theta_\mathrm{solar}} \sqrt{\Delta m^2_\mathrm{atm}} 
\lesssim | \langle m \rangle | \lesssim \sqrt{\Delta m^2_\mathrm{atm}} \,,
\end{equation}
analogous to Eq.(\ref{Aineq}) with $\Delta m^2_\mathrm{LSND}$
replaced by $\Delta m^2_\mathrm{atm}$.

\section{Summary}

Now we want to summarize our discussion of the effective Majorana mass
$|\langle m \rangle|$ relevant in $(\beta\beta)_{0\nu}$ decay and
present some numerical estimates. Using
input from existing data on neutrino oscillations and under the
assumption of small $m_1$, the following approximate bounds on (or values of) 
$| \langle m \rangle |$ are obtained. The concrete results depend on
the nature of the solution of the solar neutrino problem.

\begin{figure}
\epsfig{file=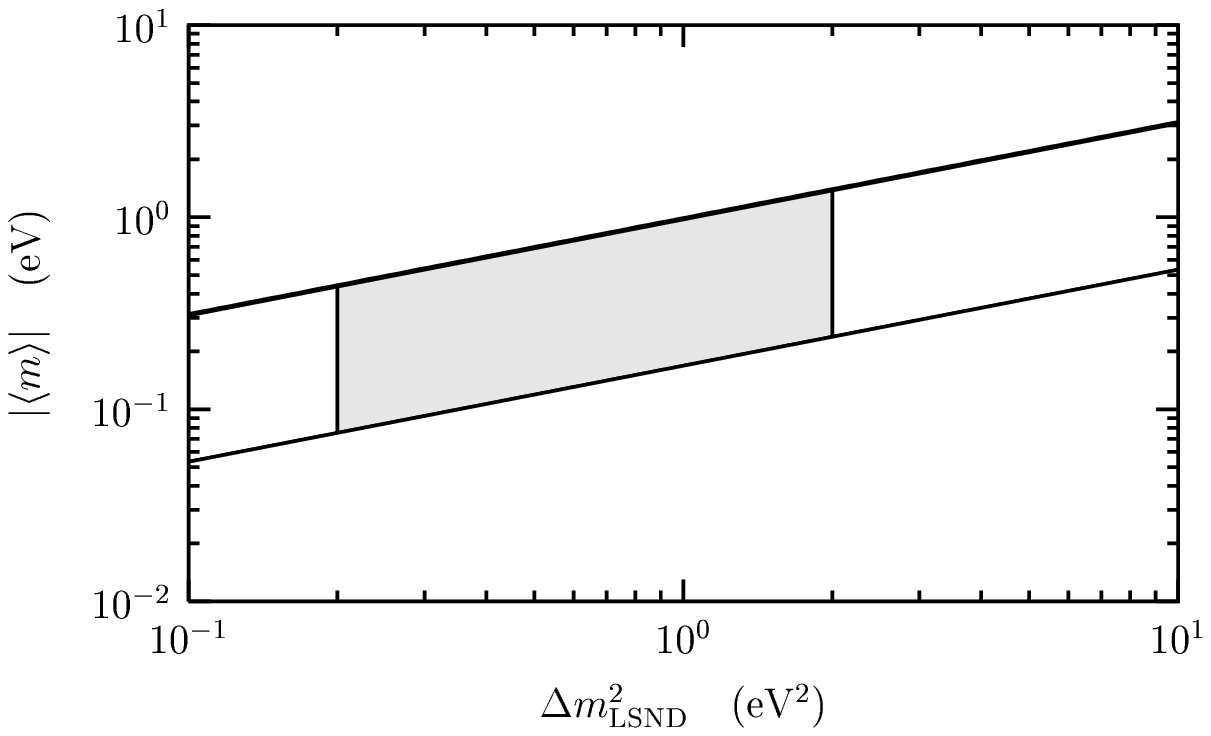,width=0.9\textwidth}
\caption{\label{bb4}Four neutrinos in Scheme A: 
The shaded area shows the possible values of 
the effective Majorana mass $| \langle m \rangle |$ in the range of 
$\Delta m^2_\mathrm{LSND}$ for the case of the LMA MSW solution of the
solar neutrino problem.}
\end{figure}

\begin{figure}
\epsfig{file=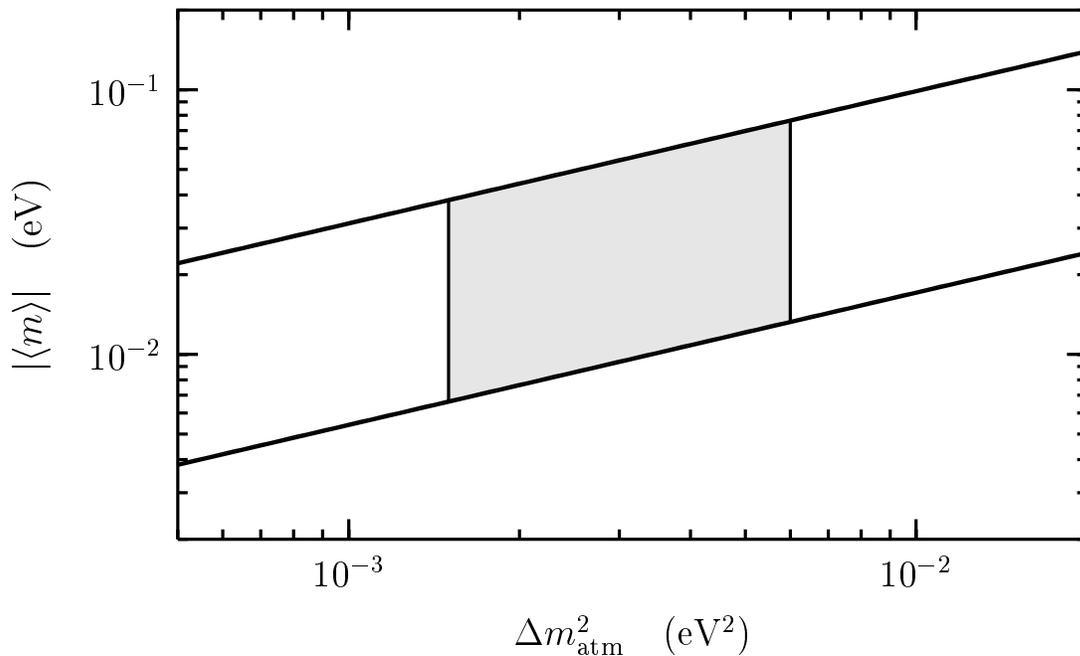,width=0.9\textwidth}
\caption{\label{bb3}Three neutrinos with reversed mass hierarchy: 
The shaded area shows the possible values of 
the effective Majorana mass $| \langle m \rangle |$ in the range of 
$\Delta m^2_\mathrm{atm}$ for the case of the LMA MSW solution of the
solar neutrino problem.}
\end{figure}

\begin{itemize}
\item[$\diamondsuit$] \textbf{4-neutrino schemes}
\begin{itemize}
\item[$\Box$] \textbf{Scheme A:}
$$
\sqrt{1 - \sin^2 2\theta_\mathrm{solar}} \sqrt{\Delta m^2_\mathrm{LSND}} 
\lesssim | \langle m \rangle | \lesssim \sqrt{\Delta m^2_\mathrm{LSND}} 
$$
If the SMA solution of the solar neutrino problem is the correct one, then
$| \langle m \rangle | \simeq \sqrt{\Delta m^2_\mathrm{LSND}}$ 
with 
$$
0.5 \; \mbox{eV} \lesssim
\sqrt{\Delta m^2_\mathrm{LSND}}
\lesssim 1.4 \; \mbox{eV}
$$
at 90\% CL \cite{LSND}.
Concerning the LMA solutions,
the lower bound on $| \langle m \rangle |$ depends strongly 
on the upper bound on $\sin^2 2\theta_\mathrm{solar}$.
In Ref. \cite{GG99} it was found for the LMA MSW solution 
in the combined analysis of all solar neutrino data that
$\sin^2 2\theta_\mathrm{solar} \leq 0.97$ at 90\% CL 
(also at 99\% CL)\footnote{We are
indebted to M.C. Gonzalez-Garcia for providing us with this number.}. 
Using this value, we have plotted in Fig.~\ref{bb4} the region defined
by the above inequalities. The shaded region represents the possible 
values of $| \langle m \rangle |$ in the allowed
range of $\Delta m^2_\mathrm{LSND}$.
We can read off from Fig.~\ref{bb4} that the effective Majorana mass lies in
the range
$7 \times 10^{-2} \lesssim | \langle m \rangle | \lesssim 1.4$ eV.
Note that for the VO solution $\sin^2 2\theta_\mathrm{solar}$ could be
as large as 1 and the lower bound in Fig.~\ref{bb4} disappears in this case.
\item[$\Box$] \textbf{Scheme B:}
$$
| \langle m \rangle | \lesssim
a^0_e \sqrt{\Delta m^2_\mathrm{LSND}}
+ m_1 +
\sin^2 \theta_\mathrm{solar} 
\frac{\Delta m^2_\mathrm{solar}}{m_1 + \sqrt{m^2_1 +
\Delta m^2_\mathrm{solar}}}
$$
For the SMA MSW solution and the VO solution this inequality reduces to
$| \langle m \rangle | \lesssim
a^0_e \sqrt{\Delta m^2_\mathrm{LSND}} + m_1$. With $a^0_e$ (\ref{a0})
as a function of $\Delta m^2_\mathrm{LSND}$ one obtains 
$$
a^0_e \sqrt{\Delta m^2_\mathrm{LSND}} \lesssim 2 \times 10^{-2} \; 
\mbox{eV}
$$
in the allowed range of $\Delta m^2_\mathrm{LSND}$ \cite{BG-WIN99}.
In the case of the LMA MSW solution also the third term in the above
inequality contributes. Using the results of Ref. \cite{GG99} 
one gets approximately
$$
\sin^2 \theta_\mathrm{solar} \sqrt{\Delta m^2_\mathrm{solar}}
\lesssim 0.3 \times 10^{-2} \; \mbox{eV}
\quad \mbox{(LMA MSW)}
$$
from the 90\% CL plot of the combined analysis. For 
$m_1 \sim \sqrt{\Delta m^2_\mathrm{solar}}$ the quantity 
$| \langle m \rangle |_2$ is, therefore, of the order of $10^{-2}$ eV.
\end{itemize}
\item[$\nabla$] \textbf{3-neutrino schemes}
\begin{itemize}
\item[$\triangle$] \textbf{Neutrino mass hierarchy:}
$$
| \langle m \rangle | \lesssim |U_{e3}|^2 \sqrt{\Delta m^2_\mathrm{atm}}
+ m_1 +
\sin^2 \theta_\mathrm{solar} 
\frac{\Delta m^2_\mathrm{solar}}{m_1 + \sqrt{m^2_1 +
\Delta m^2_\mathrm{solar}}}
$$
The term with $\sin^2 \theta_\mathrm{solar}$ is estimated as for
Scheme B. From the consideration in Subsection \ref{3I} it is clear
that the first term in this inequality cannot be larger than about
$10^{-2}$. A numerical evaluation of this term shows that 
\cite{BG-WIN99,petcov}
$$
|U_{e3}|^2 \sqrt{\Delta m^2_\mathrm{atm}} \lesssim 0.6 \times
10^{-2} \; \mbox{eV}.
$$
Thus, with $m_1 = 0$
we have $| \langle m \rangle | \lesssim 0.6 \times 10^{-2}$ eV 
in the case of  the VO or SMA MSW solution, while for the LMA MSW solution 
one finds 
$|\langle m \rangle | \lesssim 0.9 \times 10^{-2}$ eV.
\item[$\triangle$] \textbf{Reversed hierarchy:}
$$
\sqrt{1 - \sin^2 2\theta_\mathrm{solar}} \sqrt{\Delta m^2_\mathrm{atm}} 
\lesssim | \langle m \rangle | \lesssim \sqrt{\Delta m^2_\mathrm{atm}} 
$$
For the SMA MSW solution we have 
$| \langle m \rangle | \simeq \sqrt{\Delta m^2_\mathrm{atm}}$ with the
range of
$\sqrt{\Delta m^2_\mathrm{atm}}$ given approximately by \cite{SK-atm}
$$
0.03 \; \mbox{eV} \lesssim \sqrt{\Delta m^2_\mathrm{atm}}
\lesssim 0.1 \; \mbox{eV}. 
$$
For the LMA solutions the remarks for
Scheme A are valid. The possible values of $| \langle m \rangle |$ 
in the case of the LMA MSW solution are
shown by the shaded area in Fig.~\ref{bb3}, from where we get the range
$6 \times 10^{-3} \lesssim | \langle m \rangle | \lesssim 0.1$ eV.
\end{itemize}
\end{itemize}

In conclusion,
assuming $m_1 \ll m_4$ in the four-neutrino scenarios,
it is possible to have
$| \langle m \rangle |$ as large as $\sim 1$ eV in Scheme A.
In the three-neutrino scheme 
with two quasi-degenerate neutrinos and reversed hierarchy
($m_1 \ll m_2 \simeq m_3$),
$| \langle m \rangle |$ could be as large as
$\sim 0.1$ eV. However, in four-neutrino Scheme B and in the
hierarchical three-neutrino scheme, the effective
Majorana mass $| \langle m \rangle |$ is strongly suppressed, with
bounds of order $10^{-2}$ eV.
If in future
$(\beta\beta)_{0\nu}$ experiments it is found that
$| \langle m \rangle | \gg 10^{-2}$ eV it would mean that Scheme B and
the mass hierarchy with three neutrinos are excluded, or that
$(\beta\beta)_{0\nu}$ decay proceeds via
other mechanisms \cite{Mohapatra-98},
not involving the effective Majorana mass (\ref{m}). 
We would like to remind the reader that, in addition to
Scheme A, one can have
$| \langle m \rangle | > 0.1$ eV also in a three neutrino
mixing scheme with
three quasi-degenerate neutrinos \cite{Petcov-Smirnov-94,barger99,%
Vissani-bb}.

The results presented
here demonstrate that $(\beta\beta)_{0\nu}$ decay experiments are not
only important in the context of revealing the Dirac or Majorana nature of
neutrinos but also for the determination of the character of
the neutrino mass spectrum.

\acknowledgments
S.M.B. thanks the Institute for Theoretical Physics of
the University of Vienna for its hospitality and support. 
Furthermore, S.M.B.,
W.G. and S.T.P. are grateful to the CERN Theory Division for being invited
to ``Neutrino Summer '99'' during which part of this work was performed.
We also like to thank M.C. Gonzalez-Garcia and P. Krastev for discussions.
The work of S.T.P. was supported in part by the EC grant CT960090 and by
Grant PH-510 from the Bulgarian Science Foundation. 
B.K. and S.T.P. thank the Institute for Nuclear Theory at the University of
Washington for its hospitality during the completion of this work.

\end{document}